\documentclass[aps,prb,twocolumn,superscriptaddress,amsmath]{revtex4} 

\usepackage{graphicx}
\usepackage{amsfonts}
\usepackage{bbold}
\usepackage{color}

\begin{document} 

\title{Constraints on conductances for Y-junction of quantum wires}
\author{D.N. Aristov}
\affiliation{Petersburg Nuclear Physics Institute, Gatchina 188300, Russia}
\affiliation{ Institut f\"ur Nanotechnologie, Karlsruhe Institute of Technology,
 76021 Karlsruhe, Germany } 
\affiliation{Department of Physics, St.Petersburg State University, 
Ulianovskaya 1, Petrodvorets, St.Petersburg 198504, Russia}  
 \date{\today}

\begin{abstract} 
We consider the Y-junction, connecting three quantum wires, in the scattering states formalism.  
In the absence of fermionic interaction we analyze the restrictions on the values of conductance matrix, imposed 
by the unitarity of scattering $S$-matrix. Using the combination of numerical and analytical results, we describe 
the four-dimensional body of values of reduced conductance matrix. 
We show that this body touches the unit sphere 
at six points only, in accordance with Birkhoff-von Neumann theorem. 
It implies that the Abelian bosonization analysis for the vanishing interaction strength 
can be performed only in the vicinity of these six points.
\end{abstract}

\maketitle


 
\section{Introduction and model} 

Investigation of the  properties of electronic systems in reduced dimensionality is the 
subject of ongoing active studies, both theoretical and experimental. The effects of electronic 
interaction play a major role in one spatial dimension and lead to a formation of the Luttinger liquid state
\cite{GiamarchiBook}.  The idealized Luttinger liquid is best described in terms of its excitations within 
the bosonization approach. The imperfections in quantum wires, such as impurities, 
are thoroughly investigated by different methods, see \cite{Aristov2009} and references therein.  
In view of potential technological applications, it is also desirable to theoretically 
analyze junctions of several 
quantum wires.  Such analysis for the simplest variant of Y-junction connecting three wires 
was performed in a number of papers during the last  decade.  
\cite{Nayak1999,Das2004,Barnabe2005a,Oshikawa2006,Das2008,Bellazzini2009,Pugnetti2009,%
Safi2009,Agarwal2010,Aristov2010}
These works divide into two groups.  The first group considers the problem in the bosonization approach, which 
allows one to fully take into account the bulk fermionic interaction. \cite{Nayak1999,Oshikawa2006, Bellazzini2009, %
Pugnetti2009, Safi2009, Agarwal2010} 
The shortcoming of this approach is the necessity to consider the Y-junction close to certain 
values  of its transparency, as discussed below. 
Another group employs the conventional fermionic description for rather arbitrary Y-junctions, \cite{Lal2002,Das2004, 
Barnabe2005a, Aristov2010} and the drawback of these method is a finite order of fermionic interaction 
which can be taken into account.  

Recently,  a method for the fermionic description of Luttinger liquid with a single impurity was developed, 
\cite{Aristov2009} 
which performs a partial summation of fermionic diagrams in the scattering states approach. This method
allows one to obtain the exact scaling exponents known previously from the (Abelian) bosonization technique. 
It is possible to extend 
this method, formulated essentially in terms of non-Abelian bosonization, 
 for the analysis of Y-junction, \cite{unpub} and to compare its predictions with those partly available from 
numerical and bosonization studies. But before undertaking the study of the interacting case, it is necessary to 
clearly specify the restrictions on the transparency of Y-junction, 
which occur already at the level of free electrons and which 
are imposed by the basic quantum mechanical principles.  
The $3\times3$ S-matrix describing the Y-junction belongs to SU(3) group and consequences of it are sometimes  
not trivial. As we show below, the phase space for observable conductances (transparencies) is four dimensional (4D)
and the values fill the solid body of unusual shape inside the 4D unit sphere.  Importantly, only six points of this body 
lie on the surface of such a sphere, 
which implies that the Abelian bosonization analysis can be done in close vicinities of these 
points only. 
 
This paper is organized as follows. We introduce our model and basic conventions in this section below. 
The relation between the $S$-matrix and conductances is discussed in Sec.\ \ref{sec:SmatCond}. 
The main results, numerical and analytical, are presented in Sec.\ \ref{sec:values}.  We discuss our findings and  
make a comparison with previous studies in Sec.\ \ref{sec:discussion}.


\label{sec:Setup}

We consider a following setup:  three semi-infinite non-interacting wires are connected  
by the Y-junction.  The electron gas inside the wires is subject to transverse dimensional quantization, so that
only one channel per wire is available. We consider spinless fermions for simplicity. 
As usual, when interested in the response at smallest energies, 
we can linearize the electronic dispersion around the Fermi level.  
In the present case, instead of defining
right- and left-movers in each wire, it is more appropriate to speak of incoming and
outgoing waves, with respect to Y-junction at the origin.
The Hamiltonian of the system is given by
\begin{equation}
  \label{iniHam}
\mathcal{H} =  v_F \int_{-\infty}^{-0} dx \sum_{j=1}^3 ( \psi^\dagger_{j,in} i \nabla \psi_{j,in} -
\psi^\dagger_{j,out} i \nabla \psi_{j,out}  )
\end{equation}
with the Fermi velocity $v_F$ set henceforth to unity. 
We unfold the setup in the usual way, \cite{Fabrizio1995, Aristov2009} by
saying that the outgoing fermions on the negative $x$-axis are the fermions,
which have passed through the junction, at positive $|x|$, 
i.e.\ $\psi_{j,out}(-|x|) \to \psi_{j,out}(|x|) $. Thus, we
have a three-fold multiplet of right-going (chiral) fermions.  

The boundary condition at the origin is described by the scattering S-matrix. 
For elastic scattering by the central dot, the outgoing fermions
at the origin are connected to the incoming ones by the relation 
$\psi^\dagger_{k,out} = S_{km}^{\ast} \psi^\dagger_{m,in}$.  

In the scattered states representation, the right- and left- going fermionic
densities acquire the form 
\begin{equation}
\begin{aligned} \psi_{j,in} (x) &= \psi_{j} (x), \quad \psi_{j,out} (x) =
 (S.\psi)_{j} (-x) \\ \psi^\dagger_{j,in} (x) \psi_{j,in} (x) &=
\rho_j(x), \quad \psi^\dagger_{j,out} (x) \psi_{j,out} (x) =\tilde
\rho_j(-x) \end{aligned}
\end{equation}
here and below we use the notation $\tilde A = S^\dagger . A. S$.

For a multiplet of
 incoming fermions, $\Psi = (\psi_1, \psi_2, \psi_3)$, the incoming density
  $\rho_j =\Psi^\dagger .\hat \rho_j .\Psi$ is given by diagonal matrix,  
\begin{equation} 
\hat{\rho}_{1} =\begin{bmatrix} 1 & 0 &0\\ 0 & 0 &0\\ 0 & 0 &0
\end{bmatrix},  \quad
\hat{\rho}_{2} =\begin{bmatrix} 0 & 0 &0\\ 0 & 1 &0\\ 0 & 0 &0
\end{bmatrix}, \quad
\hat{\rho}_{3} =\begin{bmatrix} 0 & 0 &0\\ 0 & 0 &0\\ 0 & 0 &1
\end{bmatrix}.   
\label{rhos2lambdas}
\end{equation}
We also can write 
$\hat{\rho}_{1} =\frac{1}{2}\left( \sqrt{\frac{2}{3}}\lambda _{0}+\frac{1}{
\sqrt{3}}\lambda _{8}+\lambda _{3}\right) $, $ \hat{\rho}_{2}  
=\frac{1}{2}\left( \sqrt{\frac{2}{3}}\lambda _{0}+\frac{1}{
\sqrt{3}}\lambda _{8}-\lambda _{3}\right) $, $  
\hat{\rho}_{3}  =\frac{1}{\sqrt{6}}\lambda _{0}-\frac{1}{\sqrt{3}}\lambda _{8}$.
 Here the traceless Gell-Mann matrices, $\lambda _{j}$, with $j=1,\ldots 8$
are the generators of the SU(3) group, listed elsewhere (see, e.g., 
\cite{Tilma2002}). In addition to these, we use also the matrix $\lambda
_{0}=\sqrt{\tfrac{2}{3}}\,\mathbb{1}$, proportional to the unit
matrix, $\mathbb{1}$. We have the property $Tr[\lambda
_{j}\lambda _{k}]=2\delta _{jk}$, for $j,k=0,\ldots 8$. 

The outgoing densities are given by $\tilde \rho_j =\delta_{jk} S^\ast_{kl}
S_{km} \psi^\dagger _l \psi_m = \Psi^\dagger . S^\dagger . \hat \rho_j .S.
\Psi$, so that in the matrix representation $\hat{\tilde \rho}_j = S^\dagger
. \hat \rho_j . S $. We will mostly omit the hat sign over $\rho$ below.

\section{S-matrix and conductances}
\label{sec:SmatCond}

\subsection{Parametrization of the S-matrix}

\label{sec:introSmat}

The most general S-matrix is defined as follows 
\begin{equation}
S = \begin{pmatrix} r_{1}, & t_{12}, & t_{13}\\ t_{21} , & r_{2}, & t_{23}
\\ t_{31} & t_{32} & r_{3} \end{pmatrix}
\end{equation}
where $r_{j}$ is the reflection amplitude for wire $j$, and $t_{jk}$ is the
transmission amplitude between wires $j$ and $k$. The matrix $S$ is unitary, 
$S^\dagger S = 1$, which allows its parameterization via the exponent, $S =
\exp \left( i \sum_{j=0}^8 \theta_j \lambda_j\right)$. 
It may be more convenient to use the Euler angles parameterization of $S$. 
According to \cite{Cvetic2002},  an arbitrary $S$-matrix can be represented, up to an overall phase, as 
\begin{equation}
\begin{aligned}
S &= U e^{i \lambda_{5} \theta}  \bar U  e^{i \lambda_{8} \tau}  , \\
U & = e^{i \lambda_{3} \phi/2}e^{i \lambda_{2} \xi/2}e^{i \lambda_{3} \psi/2}, \\
 \bar U & = 
e^{i \lambda_{3}  \bar\phi/2}e^{i \lambda_{2}  \bar\xi/2}
e^{i \lambda_{3}  \bar\psi/2}.
\end{aligned}
\label{Euler1}
\end{equation}
with the range of most important parameters 
$0\le \xi\le \pi$, $0\le  \bar\xi \le \pi$, $0\le \theta \le\pi/2$, $0\le \psi \le 4\pi$.

Apparently, there is a
redundancy in the description of $S$, 
as only the densities and not the fermion amplitudes enter the
observable conductances. Below we show that there are four independent components 
of the conductance matrix, and proceed with a reduced number of parameters in the description of $S$-matrix, 
Eq.~(\ref{Euler1}). 

\subsection{Conductances}

\label{sec:Conduc}

The matrix of linear conductances $G_{jk}$ connects 
the current $I_{j}$ in wire $j$ with the electric
potentials $V_{k}$ in the leads  through the relation
 $I_{j}=\sum_{k}G_{jk}V_{k}$.  
 In linear response theory \cite{Aristov2009} the thermodynamically averaged 
 currents are given by 
 $I_{j}(x)=\int_{-\infty }^{0}dy(\langle \rho _{j}(x)\rho _{k}(y)\rangle
-\langle \tilde{\rho}_{j}(-x)\rho _{k}(y)\rangle )V_{k}$. In the static limit
and in the absence of interaction the response functions may be evaluated to
give 
\begin{equation}
G_{jk}=\delta _{jk}-Tr(\hat{\tilde{\rho}}_{j}\hat{\rho}_{k})=\delta
_{jk}-|S_{jk}|^{2}
\label{defG}
\end{equation}
One easily verifies that the charge is conserved, $\sum_{j}G_{jk}=0$ and
that applying equal voltage to all wires produces no current, 
$\sum_{k}G_{jk}=0$.

In view of these conservation laws, it is more instructive to discuss the
current response to certain combination of voltages. Let us define 
$(I_{a},I_{b},I_{0})=\hat{G}.(V_{a},V_{b},V_{0})$, with 
\begin{equation}
\begin{aligned} V_a &= (V_1 - V_2), \quad I_a  = (I_1 - I_2)/2, \\ V_b &=
(V_1 + V_2 - 2V_3)/2 \quad I_b = (I_1 + I_2 - 2I_3)/3 , \\ V_0 & = (V_1 +
V_2+V_{3})/3 , \quad I_0 = (I_1 + I_2+ I_3) , \end{aligned}
\end{equation} 
The connection between two matrices is given by  
$\hat  G= A.R^{\dagger} . G . R.A$,  with $\hat A=diag(1/\sqrt{2}, \sqrt{2/3},\sqrt{3})$ and 
\begin{equation}
R = \begin{pmatrix} 1/\sqrt{2}, & 1/\sqrt{6} , & 1/\sqrt{3} \\ - 1/\sqrt{2},
& 1/\sqrt{6} , & 1/\sqrt{3} \\ 0, & -2/\sqrt{6} , & 1/\sqrt{3} \\
\end{pmatrix},
\end{equation}
such that $R^{\dagger} R=1$. 

It follows then that the third line and the third row,
corresponding to $I_{0}$ and $V_{0}$, are identically zero, $\hat G_{j3}= \hat G_{3k}=0$.  We omit them
for clarity below, and call the upper left $2\times2$ block of $\hat G$ by the same quantity
$\hat G$.  These remaining four components are non-zero and we obtain
the reduced conductance matrix in the general form 
\begin{equation}
\begin{aligned}
\hat{G} & \equiv \begin{pmatrix} G_{aa} & G_{ab} \\ G_{ba} & G_{bb} \end{pmatrix},
\\&  =
\begin{bmatrix} \tfrac12 \left(1 - \tfrac12 Tr(\tilde\lambda_{3} \lambda_{3}) \right), &
- \tfrac 1{2\sqrt{3}} Tr(\tilde\lambda_{3} \lambda_{8}) \\ - \tfrac
1{2\sqrt{3}} Tr(\tilde\lambda_{8} \lambda_{3}) , & \tfrac23 \left( 1 - \tfrac12
Tr(\tilde\lambda_{8} \lambda_{8}) \right) \end{bmatrix} ,
\end{aligned}
 \label{defGprime}
\end{equation}
as is verified by direct calculation with Eqs.\ (\ref{rhos2lambdas}), (\ref{defG}). 

For out purposes, it is more convenient to discuss not the reduced conductance matrix $\hat G$, 
Eq.(\ref{defGprime}), but the equivalent quantity $M$ given by
 $\hat G = B.(1- M).B$, with $B=diag(1/\sqrt{2}, \sqrt{2/3})$. Explicitly, we write  
\begin{equation}
\begin{aligned} M &= 
\frac12
\begin{pmatrix}   Tr(S^{\dagger}\lambda_{3} S \lambda_{3})  , &
   Tr(S^{\dagger}\lambda_{3} S \lambda_{8}) \\  Tr(S^{\dagger}\lambda_{8} S\lambda_{3}) , &  
Tr(S^{\dagger}\lambda_{8} S \lambda_{8})    \end{pmatrix} 
, \\ & 
\equiv  \begin{pmatrix}
z -x, & y +\eta \\ y-\eta , & z+x \end{pmatrix},
\end{aligned}
\label{defMprime}
\end{equation} 

We see that, due to the charge conservation, 
the matrix of conductances contains at most four independent
components.   Further reduction in the number of independent
parameters occurs in case of additional symmetries. 
For instance, the time reversal corresponds to $\hat G \to G^{\dagger}$, so 
the unbroken time-reversal symmetry (T-symmetry) leads to $G_{ab} = G_{ba}$.  The interchange of the wires 
$1\leftrightarrow 2$ leads to a change of sign for the off-diagonal components of $\hat G$, so
the symmetry between these wires leads to $G_{ab} = G_{ba}=0$. The latter symmetry $(1\leftrightarrow 2)$ in the 
presence of the magnetic field, however, leads to conclusion $G_{ab} = -G_{ba} \neq 0$, since the parity change 
$(123)\to (213)$  should be accompanied by the change of sign of the magnetic flux piercing the Y-junction, and thus to 
transposition of $\hat G$.

These arguments are obviously applicable to a junction of $n$ wires, i.e.\ the
number of relevant physical parameters is $(n-1)^{2}$,
which is the dimension of the reduced conductance matrix. T-symmetry leads to symmetrical form $\hat G^{\dagger} = 
\hat G$, thus reducing the number of parameters to $n(n-1)/2$. 

\subsection{Reduced parametrization}

The general expression (\ref{Euler1}) contains eight Euler angles; however, the  
observable conductances (\ref{defGprime})
include only traces of products  $\lambda_{3 (8)}$ and $S^{\dagger} \lambda_{3 (8)} S$. Similarly, in the 
presence of fermionic interaction, each correction to conductances contains 
only products of $\lambda_{3 (8)}$  and  $S^{\dagger} \lambda_{3 (8)} S$. \cite{unpub}
   Since 
$\lambda_{3}$ and $\lambda_{8}$ commute with each other, the angles $\phi,  \bar \psi, \tau$ in Eq.\ (\ref{Euler1})
drop out of any observable quantity. Further, the explicit calculation shows, 
that only the combination $\psi+ \bar \phi$ enters 
our formulas and we can redefine $\psi \to \psi -  \bar \phi$ in (\ref{Euler1}) (or, equivalently, setting 
$ \bar \phi=0$ above), to eliminate $ \bar\phi$ from the subsequent analysis. (More precisely, 
 $ \bar \phi$ may be generated by the RG flow, but it does not 
enter the right-hand side of RG equations and is absent in the conductances.)
Hence, without a loss in generality, we may parametrize 
\begin{equation}
\begin{aligned}
S &= e^{i \lambda_{2} \xi/2}e^{i \lambda_{3} (\pi-\psi)/2} 
e^{i \lambda_{5} \theta} e^{i \lambda_{2}  \bar\xi/2} .
\end{aligned}
\label{Euler2}
\end{equation}

For the above matrix $M$ we have  
 \begin{equation}
\begin{aligned} M_{11}&= 
\tfrac12 \cos\xi \cos \bar\xi (1+\cos^{2} \theta)
+\sin\xi \sin  \bar \xi \cos \theta \cos\psi,  \\
M_{12}& =  
-\tfrac{\sqrt{3}}{2} \cos\xi \sin^{2} \theta ,\quad
M_{21}= 
-\tfrac{\sqrt{3}}{2} \cos  \bar \xi \sin^{2} \theta , \\
M_{22} & =
  \tfrac14(1+3\cos2 \theta)  .
\end{aligned}
\label{Mprime} 
\end{equation}
 From Eq.\ (\ref{Mprime}) it is clear that  $-1/2 \le M_{22}  \le 1$ for arbitrary $S$-matrix. 
It follows from Eq.\ (\ref{defGprime}) that the tunneling conductance $0\le G_{bb} \le 1$, as it should be.
 
 In the next section we consider two important particular cases.  
First is the time-reversal symmetrical case, when we should have $\eta=0$ in (\ref{defMprime}). 
Another case, partly considered in 
\cite{Bellazzini2007,Das2008,Hou2008,Oshikawa2006} corresponds to symmetry between the wires 1 and 2 and 
the magnetic flux piercing the Y-junction, which reads as $y=0$ in Eq.\  (\ref{defMprime}).

It should be noticed, that the correspondence between the 
arbitrary S-matrix of the form (\ref{Euler2}) and the underlying 
microscopic Hamiltonian is not simple 
beyond the lowest-order Born approximation  \cite{Aristov2009}. 
It means that the effective low-energy description for a particular microscopic setup  
can produce further restrictions on $S$, which add to those stemming from the above general symmetries.  

\section{Allowed values of conductance}
\label{sec:values}

\subsection{T-symmetric arbitrary Y-junction}  

\begin{figure} 
\includegraphics[width=5.5cm, trim = 0.0in 0.0in 0.0in 0.0in, clip=true]{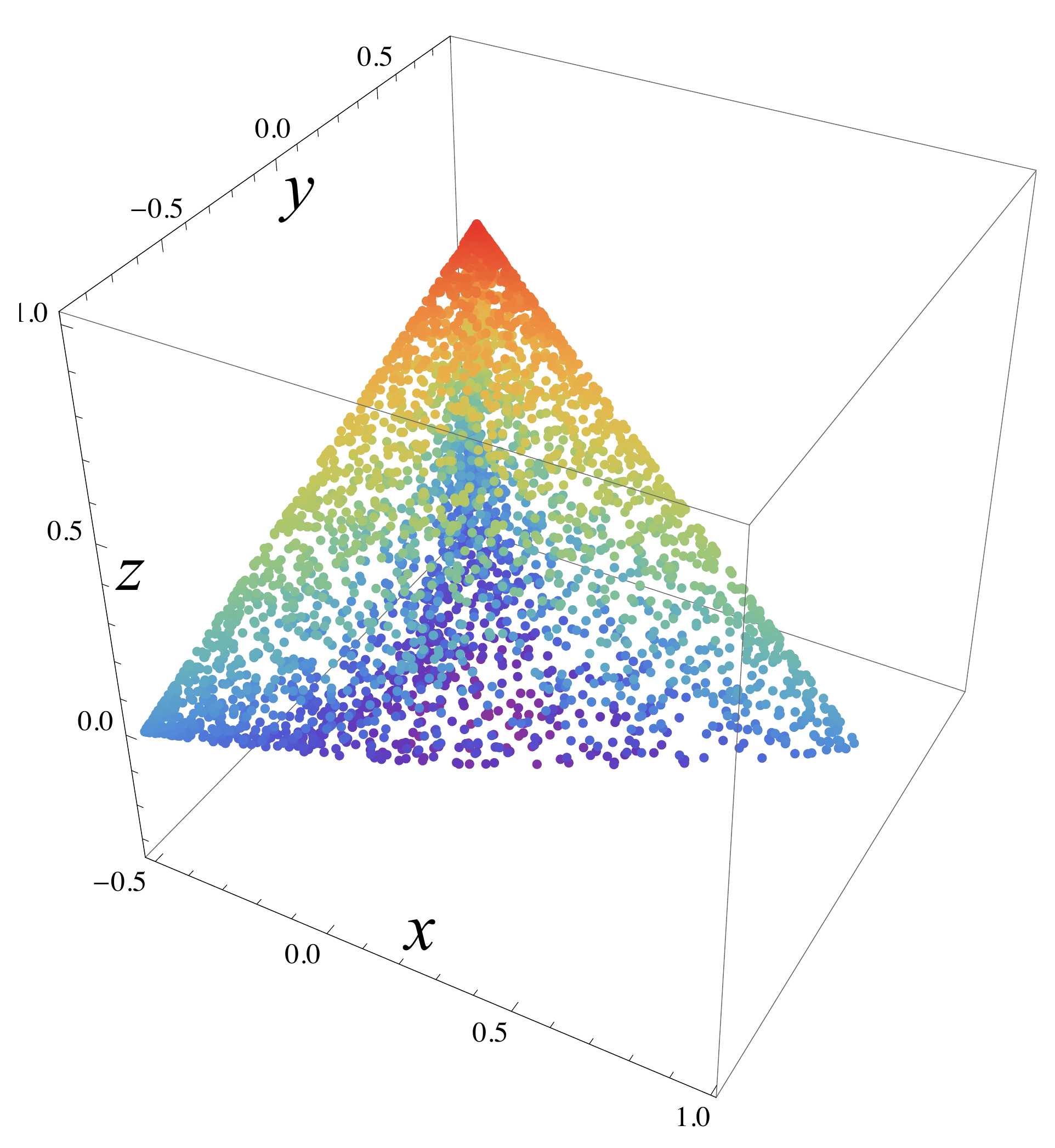}
\includegraphics[width=4.8cm]{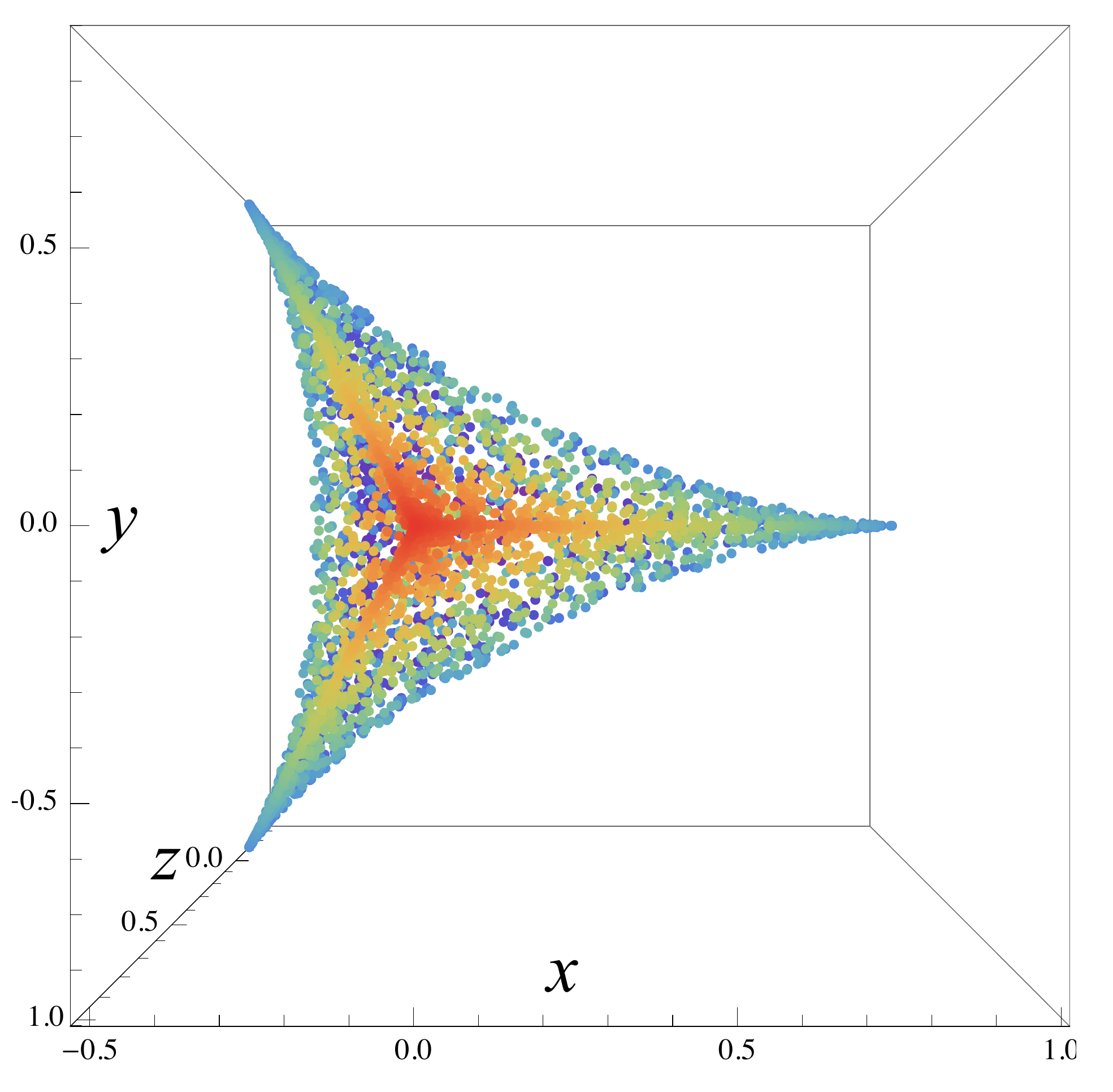}
\includegraphics[width=5cm]{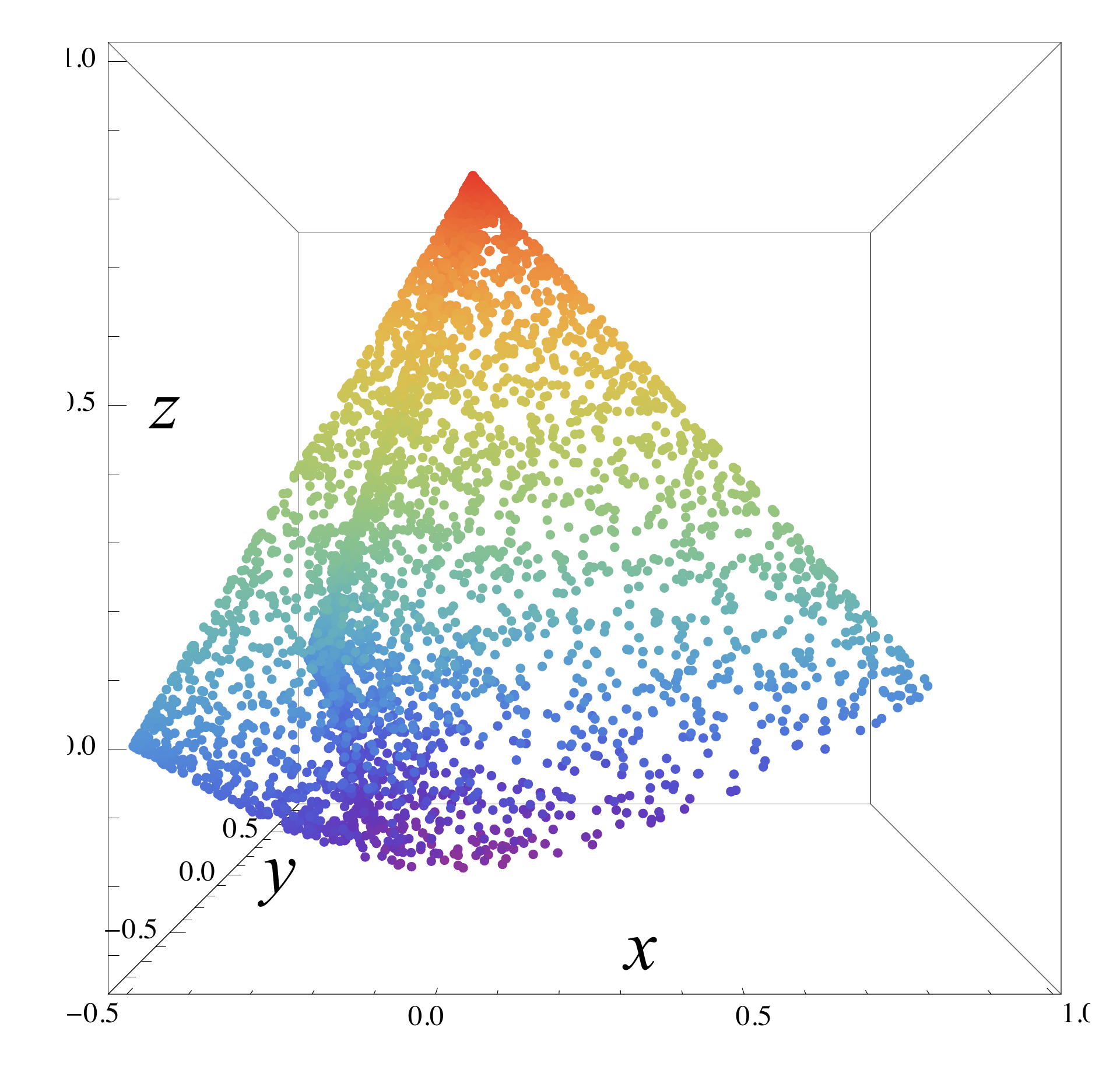}
\caption{(Color online) 
 Three views of the 
 three-dimensional body, representing all possible values of the ``conductance'' matrix $M$ in the situation 
 with the unbroken  time-reversal symmetry. 
 \label{fig:tetrahedron} } 
\end{figure}

We showed that the values of conductance are uniquely defined by the matrix $M$.
Let us discuss the case  $\eta=0$ in (\ref{defMprime}), which corresponds to time-reversal symmetry and to
 $\xi= \bar \xi$ in  (\ref{Mprime}). 
Let us take an arbitrary $S$-matrix,  given by a random realization of $\psi, \theta, \xi$ in  (\ref{Mprime}), 
and for each matrix numerically calculate three numbers $(x,y,z)$ as defined by  (\ref{defMprime}).  
The possible values 
$(x,y,z)$ do not fill a cube or a sphere, but rather form a peculiar tetrahedron, which is shown in Fig.~\ref{fig:tetrahedron}.

\begin{figure} 
\includegraphics[width=5cm, trim = 0.0in 0.0in 0.0in 0.0in, clip=true]{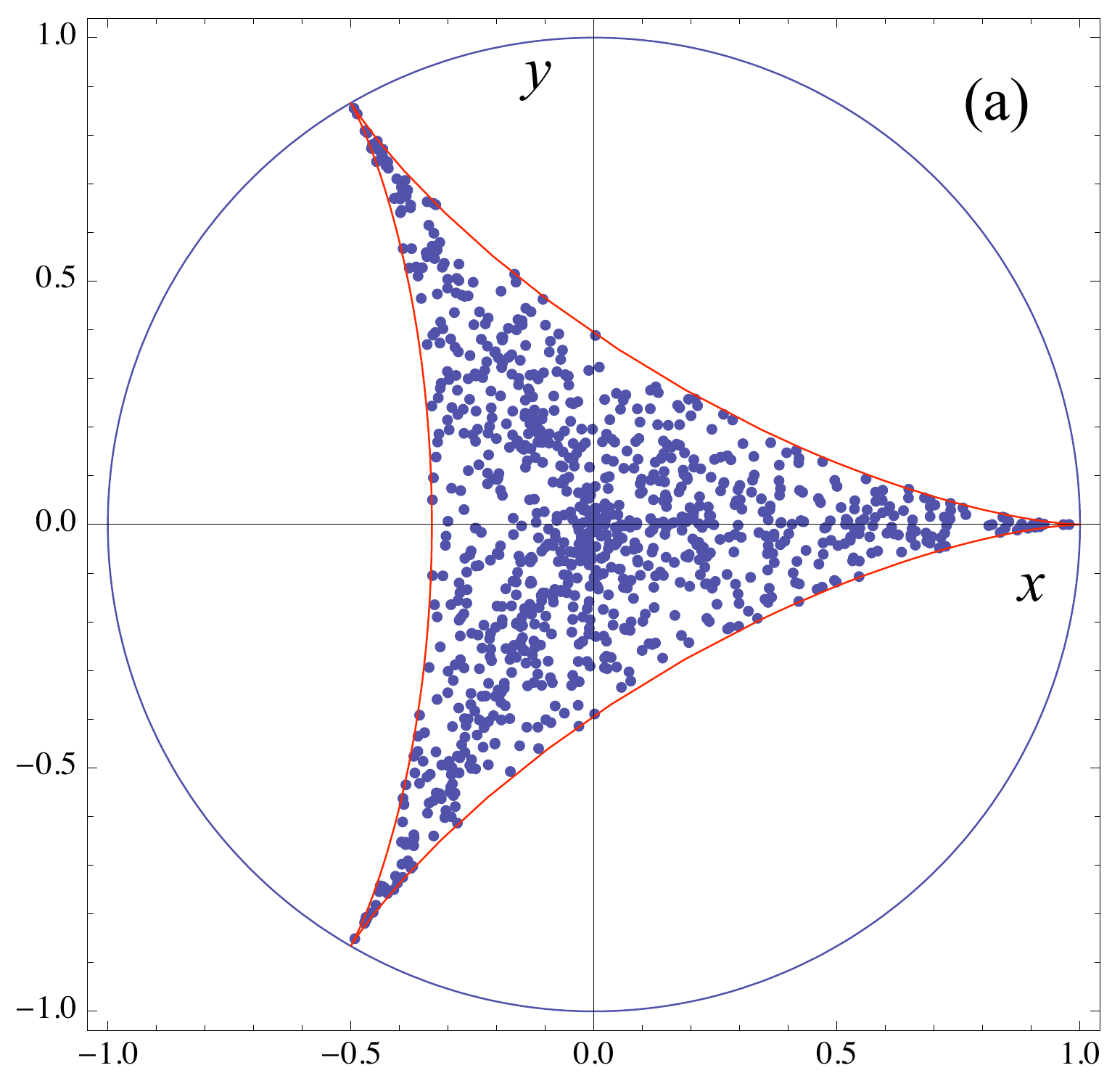}
\includegraphics[width=5cm]{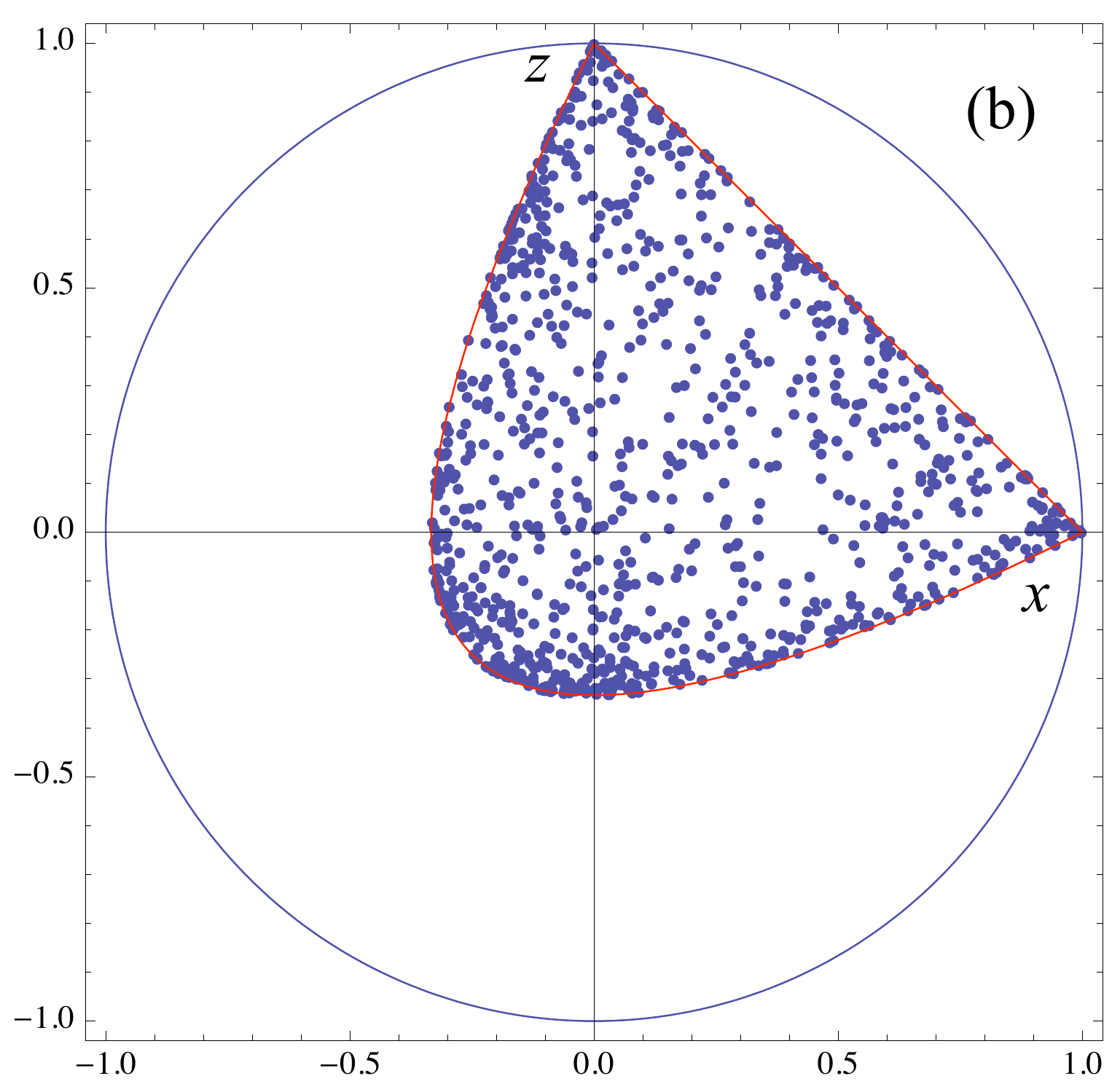}
\includegraphics[width=5cm]{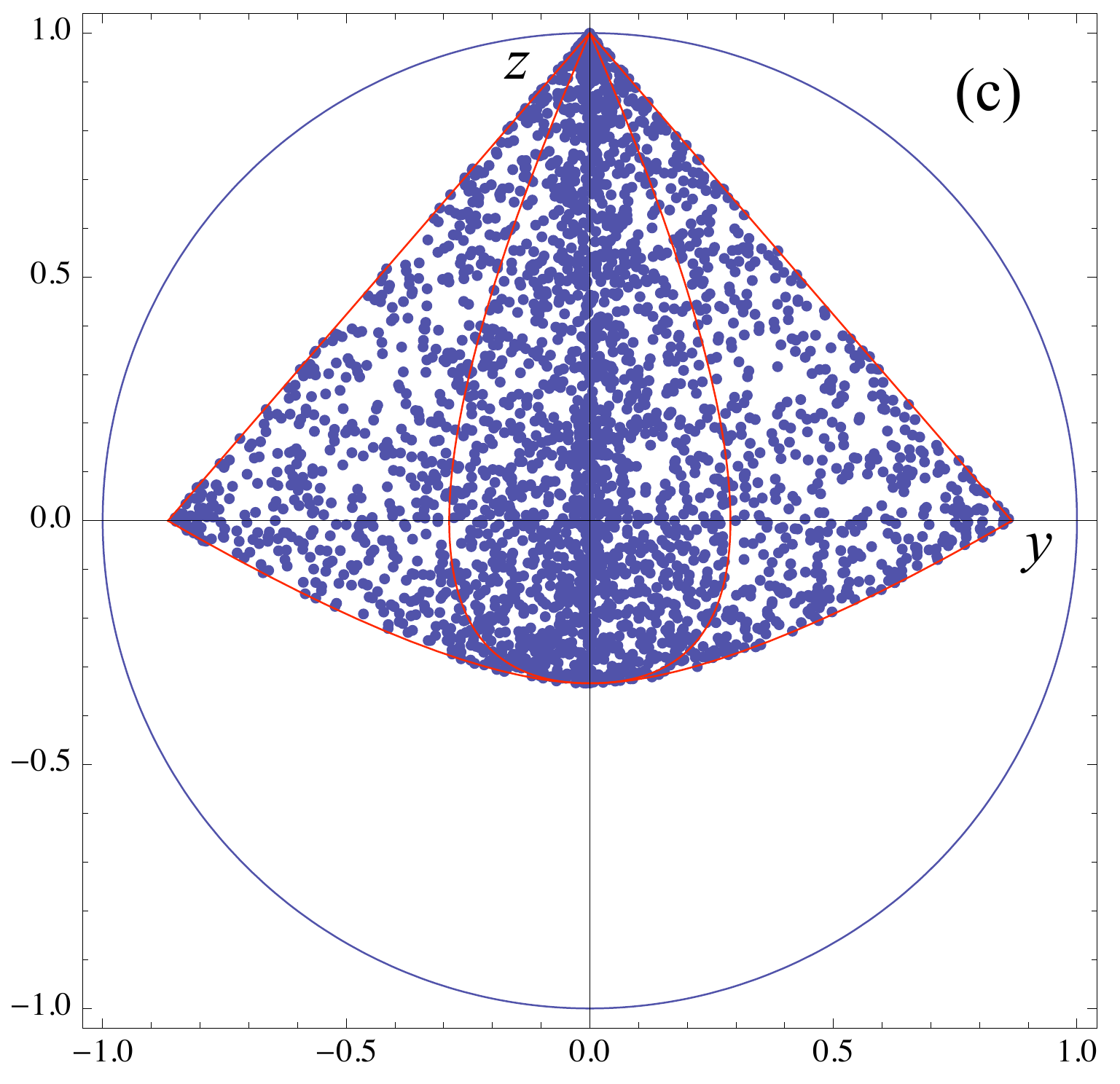}
\caption{(Color online) 
(a) A top projection of the ``tetrahedron'', showing the deltoid curve, denoted by a red line. 
The circumscribed circle is also shown. 
(b) The section $y=0$, corresponding to the symmetry between first and second wire; the bounding curve consist
of a parabola and a straight line. 
(c) The projection onto the $(x,z)$ plane; one can see parabolic and straight lines, bounding the body, 
similarly to the cross section $y=0$,  and shown by red lines. 
 \label{fig:deltoid} } 
\end{figure}

The projection of this body onto the $(x,y)$ plane is shown in Fig.~ \ref{fig:deltoid}a. 
The limiting curve, bounding the possible values $(x,y)$, is called deltoid curve and 
is given parametrically by 
\begin{equation}
\begin{aligned} x &=   (2 \cos t +\cos 2 t)/3  , 
\quad
y = (2 \sin t - \sin 2t)/3,
\end{aligned}
\label{eq:deltoid}
\end{equation}
with $t\in(0,2\pi)$.

Let us prove, that the shape in Fig.\  \ref{fig:deltoid}a  is a deltoid. 
 Introduce the notation $\tau = \cos \theta$, $u=\cos\xi$. 
Clearly, for fixed $\tau, u$ the boundary is reached at $\cos\psi=\pm1$. In view of the symmetry ($\psi \to \pi-\psi$,
$\theta\to \pi-\theta$), we may choose $\cos\psi=-1$ and 
seek the extrema of $M'$ for both signs of $\tau$. 
We have 
\begin{equation}
\begin{aligned}   
x & = - \tfrac 14 (1 + \tau) (1 + u^2 + \tau (-3 + u^2)), \\
 y& =  \tfrac{\sqrt{3}}2 (-1 + \tau^2) u .
\end{aligned} 
\label{deriv-deltoid}
\end{equation}
We find the extremum curve by 
i) fixing the direction $d(x/y)=0$ which gives the condition between $du$ and $d\tau$ 
and 
ii) expressing $du$ via $d\tau$ from this condition, we find the maximum of $x$, (simultaneously of $y$).    
It produces the parametric description of the limiting curve 
 \begin{equation}
\begin{aligned}   
x & = \tfrac 12 (3\tau^{2} -1) , 
\quad y = \pm \tfrac12 (-1 + \tau) \sqrt{3+6\tau-9\tau^{2}} ,
\end{aligned} 
\label{param-deltoid}
\end{equation}
with the range of $\tau\in (-1/3,1)$, defined by positiveness of the square root in (\ref{param-deltoid}). 
This curve coincides with (\ref{eq:deltoid}) after substitution $\tau = (1+2 \cos t)/3$.  
Importantly, the above relation $x = \tfrac 12 (3\tau^{2} -1)$ means that  $z=0$ (i.e. $M_{11}=M_{22}$) 
for the limiting curve.  


The cross section $y=0$ is shown in Fig.\ref{fig:deltoid}b. The limiting curves are straight line $x+z=1$ (corresponding to zero conductance $G_{bb}$), and a parabola, given by 
\begin{equation}
\begin{aligned} x &=(3 t - 1) (t + 1)/4 ,
\quad
z =  (3 t + 1) (t - 1)/4 \;.
\end{aligned}
\end{equation} 
The lowest point $x=y=0$, $z=-1/3$ corresponds to maximum transparency of $Y$-junction, discussed below. 

The projection onto the $(x,z)$ plane is shown in Fig.\ref{fig:deltoid}c. This projection in fact corresponds to two 
cross-sections, delimited by parabola and straight line, and rotated by an angle $2\pi/3$ in three-dimensional  space. 
The resulting contraction by a factor of $\sin(2\pi/3) = \sqrt{3}/2$ along the $x$-axis 
is taken into account,  when drawing limiting curves here.

Notice that there are only four points of ``tetrahedron'' which lie on the surface of the unit sphere 
in Fig.\ \ref{fig:tetrahedron}. 
They are $N=(0,0,1)$, $A_{3}=(1,0,0)$, $A_{1,2}=(-1/2,\pm \sqrt{3}/2,0)$ in coordinates $(x,y,z)$ and 
correspond to three detached wires, $N$, and (three times by) one detached wire, $A_{j}$, respectively. 
The related matrix $|S_{jk}|^{2}$ in (\ref{defG}) in cases $N, A_{1}, A_{2}, A_{3}$ has the form of permutation matrix, 
Eq.\ (\ref{permutmat}) below, with unity value for $p_{0},  p_{1}, p_{2}, p_{3}$, respectively. 

\subsection{Magnetic flux and symmetry between two wires}
 
In case of magnetic flux piercing the Y-junction and the additional symmetry between the wires 1 and 2, we have 
the relation  $y=0$ in  in Eq.\  (\ref{defMprime}). At the formal level, however, this case is fully analogous to the one 
considered above. Indeed, the previous case corresponded to $ \bar \xi=\xi$ and the present case 
reads $ \bar \xi=\pi - \xi$. This difference amounts to a change $(x,y,z) \to (z,\eta,x)$, 
which means that the Fig.\ \ref{fig:tetrahedron} and Fig.\  \ref{fig:deltoid} having the same form in new axes. 

The equations (\ref{deriv-deltoid}), (\ref{param-deltoid}) are now valid for $z, \eta$, and the limiting 
curve, bounding the values of conductance matrix, is again deltoid. Further, it happens at $x=0$, i.e. for the totally 
symmetric situation with respect to permutation of all three wires. Such particular case of total symmetry was considered 
earlier for interacting fermions in \cite{Nayak1999, Oshikawa2006}.
 
Similarly, there are four points which lie on the surface of the unit sphere in coordinates $ (z,\eta,x)$. They are the previous points $N, A_{3}$ and new ones $\chi_{\pm}  =(-1/2,\pm \sqrt{3}/2,0)$. The latter points are the chiral fixed points 
corresponding to currents flowing from the wire $j$ to $j\mp1$ and discussed in  
\cite{Oshikawa2006}.

\subsection{Geometrical interpretation}

In this subsection we discuss the restrictions for the elements $M_{jk}$, which stem solely from the 
unitarity of S-matrix and do not involve particular properties of SU(3) group.  
 We represent Eq. (\ref{defMprime1}) in the form   
\begin{equation}
\begin{aligned} M&= \begin{pmatrix} \langle \tilde \lambda_{3} |
\lambda_{3} \rangle, & \langle \tilde\lambda_{3} | \lambda_{8} \rangle \\
\langle \tilde\lambda_{8} | \lambda_{3} \rangle, & \langle \tilde
\lambda_{8} | \lambda_{8} \rangle \end{pmatrix} 
\equiv \begin{pmatrix}
a \cos\alpha , & b \cos \beta \\ a \sin\alpha, & b \sin \beta \end{pmatrix},
\end{aligned}
\label{defMprime1}
\end{equation}
here the shorthand notation $\langle x | y\rangle = \tfrac 12 Tr( x . y)$ was introduced. 

It is seen, that the columns of $M$ are projections of $\lambda_{3}$
and $\lambda_{8}$ onto the hyperplane spanned by $\tilde \lambda_{3}$ and 
$\tilde \lambda_{8}$. We can think of the vectors $\vec a = \tilde
\lambda_{3} \langle \tilde \lambda_{3} | \lambda_{3} \rangle + \tilde
\lambda_{8} \langle \tilde \lambda_{8} | \lambda_{3} \rangle $ and $\vec b =
\tilde \lambda_{3} \langle \tilde \lambda_{3} | \lambda_{8} \rangle + \tilde
\lambda_{8} \langle \tilde \lambda_{8} | \lambda_{8} \rangle$. Each of the sets 
$\{\tilde \lambda_{j} \}$ and $\{\lambda_{j} \}$ with $j=0,\ldots 8$ forms the orthonormal basis 
for Hermitian matrices with the scalar product $\langle \cdot | \cdot \rangle$. 
Therefore we have $a^{2} = \sum
_{j=3,8} |\langle \lambda_{3} | \tilde\lambda_{j}\rangle|^{2} \leq \sum
_{j=1}^{8} |\langle \lambda_{3} | \tilde\lambda_{j}\rangle|^{2} =1$ and
similarly $b^{2}\leq1$. Notice that these inequalities follow from the
unitarity of $S$, i.e.\ charge conservation. 

Further,  let us consider a unit
vector $\vec v = \cos \gamma\, \lambda_{3} + \sin \gamma\, \lambda_{8}$ in the
hyperplane spanned by $\lambda_{3}$ and $\lambda_{8}$. The projection of the
circle $\gamma\in (0,2\pi)$ onto the plane $\{ \tilde \lambda_{3}, \tilde
\lambda_{8} \} $ is given by $\vec v_{P} = \cos \gamma\, \vec a + \sin \gamma\,
\vec b$ and should have the form of the ellipse. A simple calculation shows
that the principal axes of this ellipse are given by 
\begin{equation}
\cos^{2} \zeta_{\pm} = \tfrac12 ({a^{2}+b^{2}}) \pm \tfrac12 \sqrt{%
(a^{2}-b^{2})^{2} + 4(\vec a.\vec b)^{2} }, 
\end{equation}
here $\vec a.\vec b = a b \cos(\alpha- \beta )$ 
Evidently, $\zeta_{-}$ is the \emph{dihedral angle} between the hyperplanes 
$\{ \tilde \lambda_{3}, \tilde \lambda_{8} \} $ and $\{ \lambda_{3},
\lambda_{8} \} $. We also have an obvious geometrical inequality,
$ \cos^{2} \zeta_{+} \le 1$,
which can be written in the form 
\begin{equation}
a^{2} + b^{2} \leq 1+ a^{2} b^{2 } \sin^{2}(\alpha-\beta)
\label{ineq:conduc}
\end{equation}
and amounts to the statement that the eigenvalues 
of the non-negative matrix
$  M^{\dagger} .M =   \begin{pmatrix}
a ^{2} , &\vec a.\vec b  \\ \vec a.\vec b , & b^{2} \end{pmatrix},$ 
are less or equal to unity,  $\| M^{\dagger} M\| \leq1$.

\subsection{Birkhoff-von Neumann theorem}

Introducing $ \bar M_{ij} = |S_{ij}|^{2}$ in Eq.\ (\ref{defG}), 
we write $G_{ij} = \delta_{ij} - \bar  M_{ij}$. 
We have the property $\sum_{i} \bar  M_{ij} =\sum_{j} \bar M_{ij} =1$, 
the matrices of this form are called doubly stochastic.  
For a vector space with elements $X_{i}$ we may form a linear combination of the form $\sum_{i} c_i X_i$. If
each $c_i \ge 0$ and $\sum_i c_i = 1$, then such combination is called convex combination. 

The Birkhoff-von Neumann theorem states that an $n \times n$ matrix over $\Re$ is 
doubly stochastic if and only if it is a convex combination of permutation matrices. 

A doubly stochastic matrix is called unistochastic if its elements can be represented as squared moduli 
of elements of unitary matrix. Any unistochastic matrix is contained in 
the set of doubly stochastic matrices, but not vice versa. In our case a general doubly stochastic matrix 
is a candidate for conductance matrix, but not all such matrices have a quantum-mechanical counterpart, $S-$matrix, 
i.e. they do not belong to unistochastic set. 
The analysis of
$3\times 3$ unistochastic matrices, done in  [\onlinecite{Bengtsson2005,Zyczkowski2003}],  
is relevant to our study. 
Particularly, it was shown, that the extreme points of unistochastic set are permutation matrices.   
The doubly stochastic property represents the classical 
Kirchhoff's rule for the charge conservation, whose quantum-mechanical counterpart,  
the unitarity of $S-$matrix, 
necessarily involves the complex-valued quantities, except for extreme points and surfaces.
A special role is played by two subsets of matrices of the form (cf. Eq.(16) in \cite{Agarwal2010})

\begin{equation}
\begin{aligned} \bar M_{1} &= \begin{pmatrix} 
  p_{0}, &p_{4}, &p_{5},\\
  p_{5}, &p_{0}, &p_{4},\\
  p_{4}, &p_{5}, &p_{0}
  \end{pmatrix}, 
\quad \bar  M_{2} = \begin{pmatrix} 
  p_{1}, &p_{3}, &p_{2},\\
  p_{3}, &p_{2}, &p_{1},\\
  p_{2}, &p_{1}, &p_{3}
  \end{pmatrix} .
\end{aligned} 
\label{permutmat}
\end{equation} 
The above mentioned extreme cases correspond to $p_{i}=1$ for a given $i$, with all other   $p_{i}=0$. 
These cases correspond to the above points 
 $N$, $A_{1}$,  $A_{2}$, $A_{3}$, $\chi_{+}$, $\chi_{-}$, for $i=0,\ldots 5$, respectively.

Explicitly, we have in general case of  $M$, Eq.\ (\ref{defMprime}), 
 \begin{equation} 
\begin{aligned}  x &= p_{3} - \tfrac12 (p_{1} + p_{2}), 
\quad   
y= \tfrac{\sqrt{3}}2 (p_{1} - p_{2}),
\\
z&= p_{0} - \tfrac12( p_{4} + p_{5}),
\quad 
\eta =  \tfrac{\sqrt{3}}2(p_{4} - p_{5}),
\end{aligned} 
\label{2stoch}
\end{equation}
with $\sum_{i=0}^{5}p_{i}=1$.
 
\begin{figure} 
\includegraphics[width=5cm, trim = 0.0in 0.0in 0.0in 0.0in, clip=true]{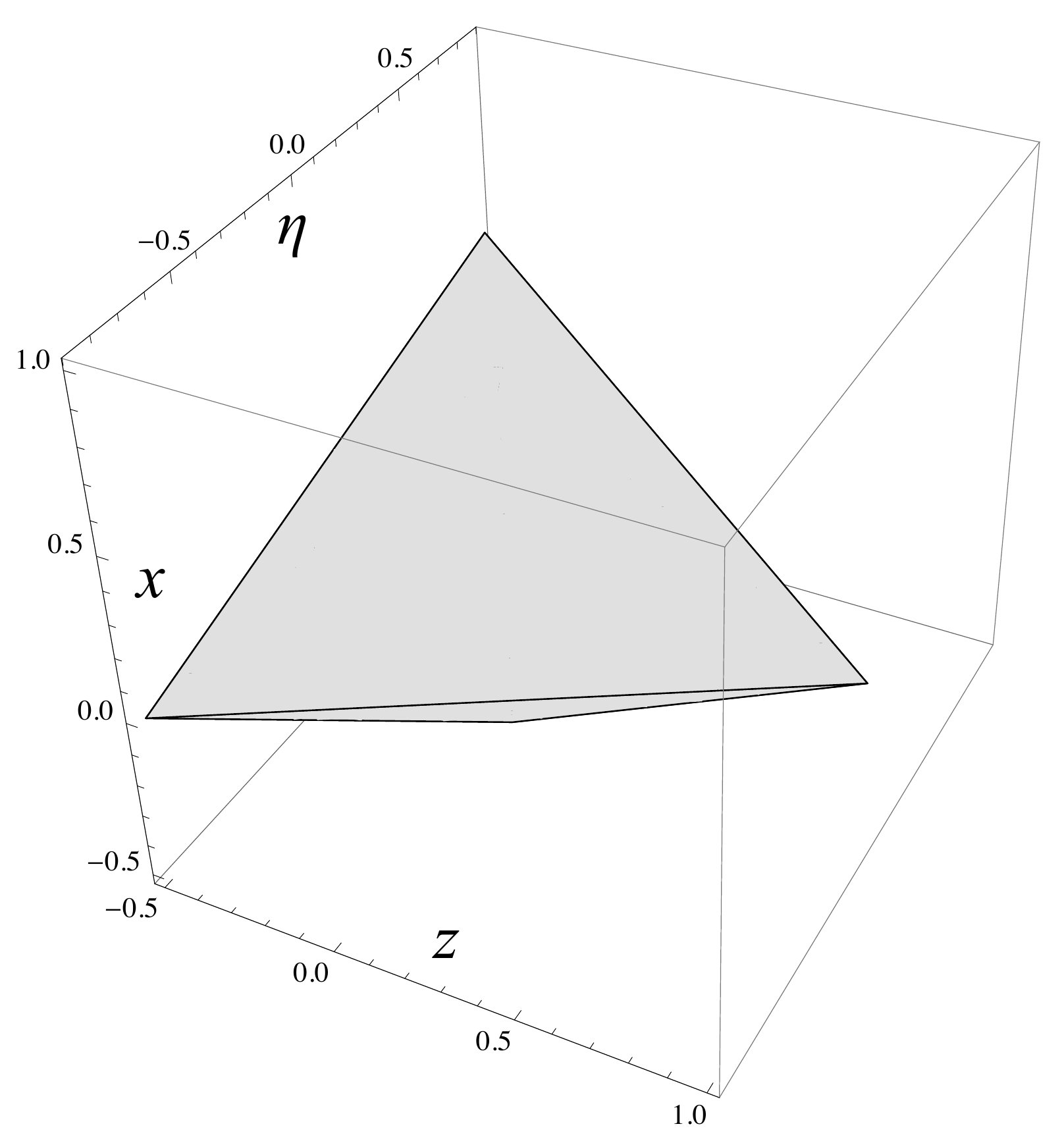}
\includegraphics[width=5cm]{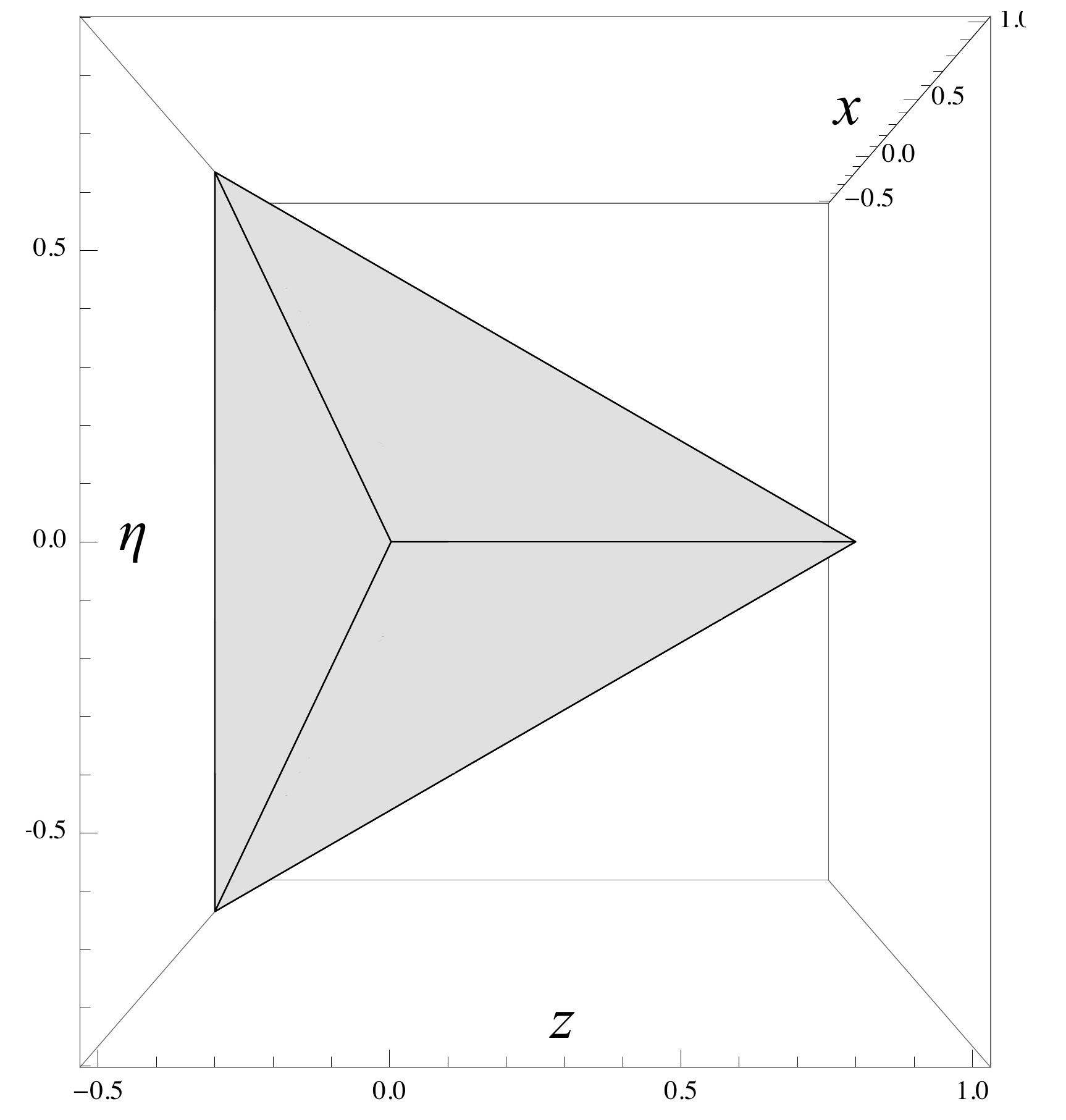}
\includegraphics[width=5cm]{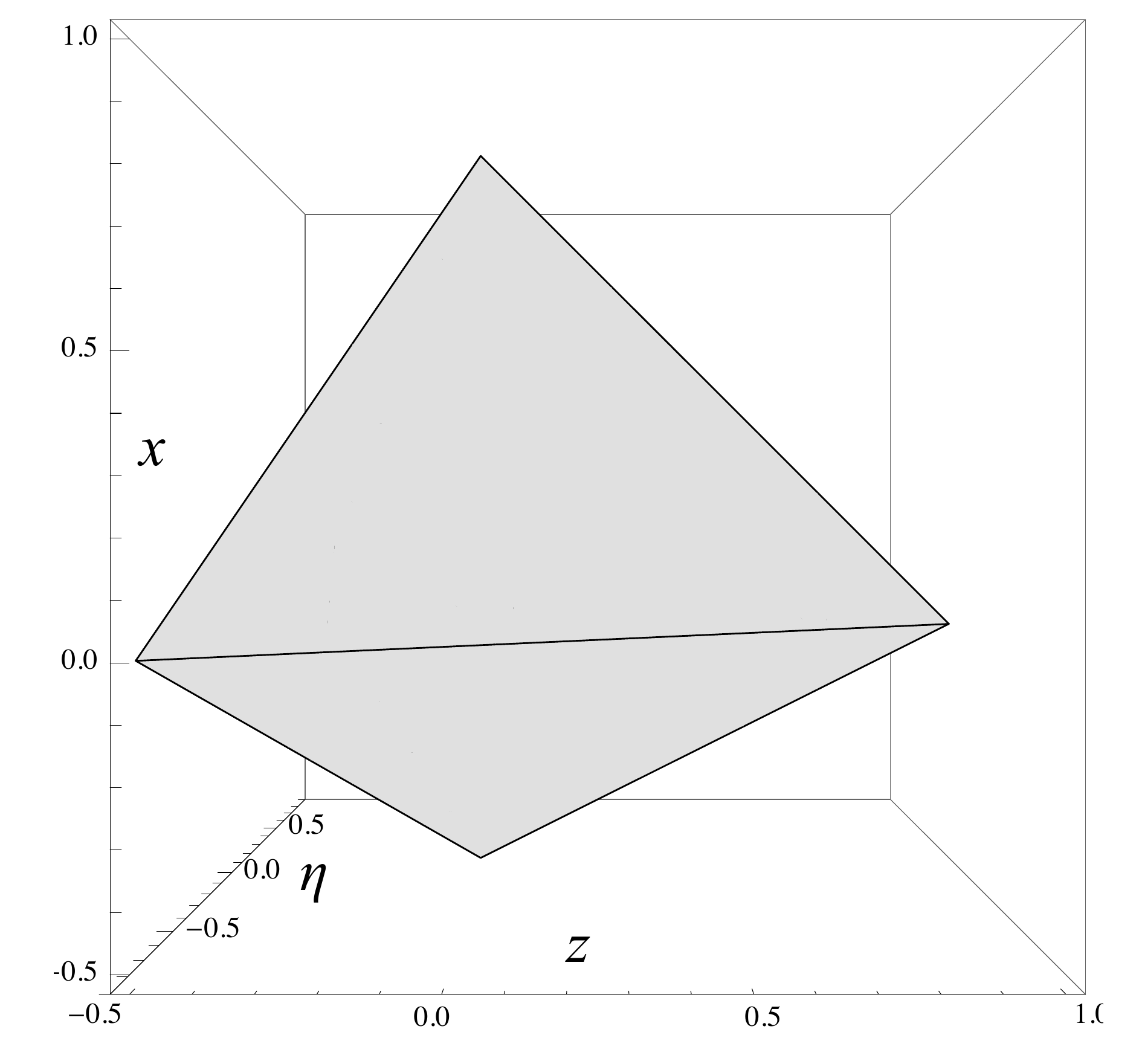}
\caption{
 Three views of bipyramid, representing all possible values of classical ``conductance'' matrix $M$ in the symmetric 
 situation ($y=0$ in Eq.\ (\ref{2stoch})) with broken  time-reversal symmetry.   
 \label{fig:bipyramid} } 
\end{figure}

The Birkhoff-von Neumann theorem does not imply a uniqueness of the convex combination. This becomes 
evident when we notice that $\bar M_{1}+ \bar M_{2} = 0$ for arbitrary $p_{0}$, if
$p_{0}=p_{4}=p_{5}=-p_{1}=-p_{2}=-p_{3}$. For further illustration of this issue, consider the 
above additional condition of unbroken time-reversal symmetry 
$\eta=0$ which reduces the freedom in choice of parameters $p_{j}$, but not completely. The allowed values 
 of  $x,y,z$ for the doubly stochastic matrix $\bar M_{1}+ \bar M_{2}$ of the above form fill the triangular bipyramid, 
 which is the ``classical'' analog of the body shown in Fig.\  \ref{fig:tetrahedron}. 
 This bipyramid is defined by its vertices $(x,y,z)$ at $(0,0,1)$, $(1,0,0)$, $(-1/2,\pm\sqrt{3}/2,0)$ and 
 $(0,0,-1/2)$.
The consideration of the case with magnetic flux, $y=0, \eta \neq 0$ is done similarly to above, 
due to obvious symmetry  of Eq.\  (\ref{2stoch}). We have the same bipyramid in coordinates $(z,\eta,x)$, as 
depicted in Fig.\ \ref{fig:bipyramid}.

Concluding this section, we make two observations. 
First, there are only six points, which lie on the surface of the unit four-dimensional sphere $(x,y,z,\eta)$. 
They are $N$, $A_{1,2,3}$, $\chi_{\pm}$ and this set is the same for doubly stochastic and unistochastic matrices. 
Second, the above inequality (\ref{ineq:conduc})  
can be cast into the form 
\begin{equation} 
 \sqrt{x^{2}+y^{2}} + \sqrt{z^{2}+\eta^{2}} \le 1 .
 \label{restr-xyz}
 \end{equation}
This condition at $\eta=0$ represents two cones. 
Three straight lines, connecting $N$ and $A_{j}$ lie on the upper cone $z=1- \sqrt{x^{2}+y^{2}}$ and 
belong to the boundary of allowed values of conductance both for doubly stochastic and unistochastic sets.
The lower cone, $z=-1+ \sqrt{x^{2}+y^{2}}$, does not touch the above ``tetrahedron'' in both variants.   


 \section{Discussion}

\label{sec:discussion}

\subsection{``Stability of the action''} 

\label{sec:stability}

Consider the energy, associated with the chiral fermionic densities in
wires. For the ring of length $l$ the energy of the
chiral current $\rho_{j}$ around the energy minimum is given by $\pi v_{F} l^{-1}
(\rho_{j} -\rho_{j}^{(0)})^{2} $, see e.g.\ \cite{Aristov2002a, Aristov2007} and references
therein. Here $\rho_{j}^{(0)}$ is discrete variable (zero mode), 
measuring the number of particles
moving at a given direction. Extending this formula to our situation, we have
the energy of the incoming and outgoing d.c.\ currents  per unit length
\begin{equation}
\begin{aligned}
E_{in} &= \pi v_{F} \sum_{j=1}^{3} \Big( l^{-1} \int_{-l}^{0}dx\, \langle
\rho_{j,in}(x) \rangle \Big)^{2}, 
\\ 
E_{out} &= \pi v_{F} \sum_{j=1}^{3} \Big(l^{-1} \int_{0}^{l}dx\, \langle
\rho_{j,out} (x)\rangle \Big)^{2}
\end{aligned}
\label{eq:energycurrent}
\end{equation}
with  $l\to \infty$. 
This formula is written for the non-interacting fermions, and with some restrictions
 it can be used in the interacting case, as well. 
Such generalization is possible, e.g., when one considers the finite region of the interaction 
close to the origin, $|x|<L$,  in  Eq.\ (\ref{eq:energycurrent}).   
In this case the contribution of the interacting region $x\in (-L,L)$ in the semi-infinite leads
to $E_{in,out}$ is negligible,
and we can adopt the single-particle description of the outgoing states and the
 free-fermion relation between the applied voltage and the incoming current, $\langle \rho_{j,in}
\rangle = \tfrac {e^{2}}{h} V_{j}$. The existence of the non-interacting leads beyond the finite region of interaction is important for the proper definition of $S$-matrix;  
it also leads to the absence of vertex corrections to conductance and the (correct) unity value of conductance of clean Luttinger liquid, as discussed at length in \cite{Aristov2009}.
Here $\langle \rho_{j,in}\rangle $ stands for  zero mode of the current, i.e.\ 
$\langle \rho_{j,in}(x)\rangle$  averaged over $x$ in (\ref{eq:energycurrent}). 
The linear response theory gives for the net current $I_{j}=\langle \rho_{j,in} \rangle - \langle
\rho_{j,out} \rangle = \tfrac {e^{2}}{h} G_{jk}V_{k} = G_{jk} \langle
\rho_{k,in} \rangle $. Therefore $\langle \rho_{j,out} \rangle = \bar  M_{jk} 
\langle \rho_{k,in} \rangle $,  with the above matrix $ \bar M_{jk} = \delta_{jk} - G_{jk} = |S_{jk}|^2 $. 
Notice that, by construction, $Tr(\bar M) \ge 0$ for any $S$.

The energies (\ref{eq:energycurrent}) represent \emph{only a part} of the energy of chiral fermions
associated with the zero modes of the system. 
Indeed, from the general inequality $(\int f g)^{2} \le (\int f^{2})(\int g^{2})$ we obtain 
$ \pi v_{F}  \Big( l^{-1} \int_{-l}^{0}dx\, \langle
\rho_{j,in}(x) \rangle \Big)^{2} \le \pi v_{F}   l^{-1} \int_{-l}^{0}dx\, \langle
\rho_{j,in}^{2}(x) \rangle \equiv E_{tot}$, where the last quantity is the total energy density in 
bosonization description.  It is assumed, that the total energy of incoming fermions is 
given by the zero modes, $E_{in} =  E_{tot}$. 

Since the scattering is elastic, and a part of the
energy $E_{in}$ after the scattering may become distributed among the
excited particle-hole pairs (with non-zero momentum in appropriate
description), one obtains the inequality $E_{in} \geq E_{out}$ which was called a 
stability condition in  \cite{Oshikawa2006}.
 This inequality means, that the eigenvalues of the non-negative matrix
  $\bar M^{\dagger} \bar M$, or equivalently  $M^{\dagger} M$, are less or equal to unity.   

We saw above, however, that the stability condition $\| M^{\dagger} M\| \leq1$
follows automatically from the unitarity of $S$. 
The bosonization approach \cite{Bellazzini2007,Das2008,Hou2008,Oshikawa2006}
does not assume the unitarity of $S$-matrix (see below), but  
rather starts from the consideration of conductance matrix, $G$, or 
 ``current splitting'' matrix, $\bar M$. 
As shown above, Eqs.\  (\ref{defMprime1}), (\ref{ineq:conduc}), the restrictions for the parameters are 
\begin{equation}
\begin{aligned} a^{2} &\leq 1 , \quad b^{2} \leq 1 , \\
\cos^{2}(\alpha-\beta) &\leq (a^{-2} -1)(b^{-2}-1) \end{aligned}
\label{restrict}
\end{equation}

The case considered in Ref.\ \cite{Oshikawa2006} reads in our notation as $a\cos\alpha =
1-\tfrac32 G_{s}$, $a\sin\alpha = \tfrac32 \Delta$, $b=a$, $\beta =
\alpha+\pi/2$. We see that the unitarity-related inequalities (\ref{restrict}) reduce in this case to 
$(1-\tfrac32 G_{s})^{2}+ (\tfrac32 \Delta)^{2} \le 1$, which correspond to  Eq.\ (11.13) in \cite{Oshikawa2006}.
At the same time, Figs.\  \ref{fig:deltoid}, \ref{fig:bipyramid} show that the bounding estimates 
(\ref{restrict}), (\ref{restr-xyz}) are too weak. 

\subsection{Abelian bosonization}

Let us further comment on the nature of the inequality  $E_{in} \geq E_{out}$  
linking it to the bosonization approach. According to scattering states formalism
described above and in Ref.\ \onlinecite{Aristov2009}, the components of outgoing density may include
non-diagonal interference terms. For example, the appearance of $\lambda_{4}$ component in the 
outgoing current $\tilde \rho_{1}$ indicates the presence of combination
$ \psi^{\dagger}_{3}\psi_{1} + h.c. $.  
Contrary to the interpretation in \cite{Agarwal2010}, the existence of such components of the density 
does not mean the power dissipation, as the scattering is elastic. The difference between the diagonal terms 
$\psi^{\dagger}_{j} \psi_{j}$ and off-diagonal terms $\psi^{\dagger}_{j}\psi_{i}$ ($i\neq j$) becomes more transparent,
if we consider a non-equilibrium situation with voltage biases $V_{j} \neq 0$. In this case we should recall the 
existence of the rapidly oscillating exponents in the wave-function of the incoming fermion, 
$\psi_{j}(x) \sim \bar\psi_{j} (x) \exp i k_{Fj} x$, with Fermi momenta $k_{Fj} = k_{F} + V_{j}/v_{F}$; 
here we extracted the ``smooth'' part of the fermion operator $\bar\psi_{j} (x)$, see e.g.\ \cite{GoNeTs}. 
We see now that the diagonal  terms $\psi^{\dagger}_{j} \psi_{j}$ in the outgoing current contain only smooth part
of the density, whereas the off-diagonal part of the density contains ``beats'' in its profile, 
$\bar \psi^{\dagger}_{j} \bar\psi_{l} \exp i ( V_{j}- V_{l}) x/v_{F}$,  cf.\ Ref.\  \cite{Urban2008}. 
Evidently, the off-diagonal terms vanish upon the averaging in (\ref{eq:energycurrent}), and this is exactly when 
the energy difference $E_{in}  - E_{out}$ occurs.  This difference 
is thus associated with the weight of the off-diagonal components of 
fermionic density, which may also include an analog of $2k_{F}$ Friedel oscillations.  

The condition of exact equality $E_{in}  = E_{out}$ indicates a possibility to fully express the outgoing currents 
in ``smooth'' components of incoming chiral densities. It means that, in the appropriate basis, the whole 
Hamiltonian is expressed entirely in smooth densities which is the essence of Abelian bosonization approach.
Notice, that it is always possible to express the kinetic part of the action in terms of smooth 
densities in each of the semi-infinite wires. The difficulty in such a description arises at the moment of imposing 
a boundary conditions, which connect the smooth densities in different semi-wires. \cite{Oshikawa2006,Hou2008}
Thus, for instance, a simpler case of the wire with a single impurity allowed a simple perturbative analysis only
in two limiting cases of a clean wire and two detached semi-wires. \cite{Kane1992} 

Evidently, the condition $E_{in}  = E_{out}$ happens at   $a=b=1$ and 
$\alpha=\beta\pm\pi/2$ in (\ref{defMprime1}).  We have two classes discussed in 
\cite{Agarwal2009,Agarwal2010,Bellazzini2009} namely 
 \begin{equation}
\begin{aligned} M _{1}&=  \begin{pmatrix}
\cos\alpha , &  \sin \alpha  \\   - \sin\alpha, & \cos \alpha \end{pmatrix}, \\ 
 M _{2}&=  \begin{pmatrix}
-\cos\alpha , &  \sin \alpha  \\    \sin\alpha, & \cos \alpha \end{pmatrix}. \\ 
\end{aligned}
\label{M12}
\end{equation}
 (the form of $M_{1}$  in our notation has a direct correspondence with Eq.\ 
(10.17) in \cite{Oshikawa2006} ).
In these cases $M$ is orthogonal, $M^{\dagger}M=1$, and one may promote the above equality 
for the averages to
the operator form, $\rho_{j,out} = \bar M_{jk} \rho_{k,in} $. Then the commutation relations 
for the chiral fields,  
$[\rho_{j,in}(x),\rho_{k,in}(y)] = \tfrac i{2\pi}  \delta_{jk}\delta'(x-y)$,  prescribed in bosonization, are 
fulfilled also for  $\rho_{j,out}$.   

In (Abelian) bosonization approach, the consistent perturbative analysis can be done only 
in a close vicinity of the RG fixed point,  which is scale invariant or conformally invariant point in 1D. 
\cite{Bellazzini2007,Das2008,Hou2008} 
The requirement for the above matrices (\ref{M12}) to represent the RG fixed point in bosonization 
leads to additional boundary conditions for three chiral densities at the Y-junction. These boundary conditions
are expressed via a set of Neumann and Dirichlet boundary conditions for appropriate bosonic fields. 
An important observation, made in Ref.\ [\onlinecite{Bellazzini2007}], is that  the trace of the matrix $\bar M$ at the fixed
point is an integer number and equal to the number of Neumann conditions imposed. In simple words, 
 $Tr(\bar M) = 1 + Tr(M)=1+2z$,  at the fixed point 
should be equal to the number of wires, detached from the $Y$ junction; it becomes rather obvious when we 
recall that $ Tr (\bar M) = |r_{1}|^{2} + |r_{2}|^{2} +  |r_{3}|^{2}$. Evidently, the number of detached wires can be 
$0, 1, 3$, but not 2.

The fixed points discussed in bosonization approach include  the above $N$ point, 
$M_{1} (\alpha=0) =  \left( \begin{smallmatrix} 1  , &  0 \\   0, & 1\end{smallmatrix} \right),$ 
for three detached wires ; the $A_{3}$ point,  
$M_{2} (\alpha=0) =  \left( \begin{smallmatrix}  -1  , &  0 \\   0, & 1\end{smallmatrix}\right),$ 
for the third wire detached, and 
also $M_{2} (\alpha=\pm2\pi/3) =  \left( \begin{smallmatrix}1/2  , &  \pm \sqrt{3}/2 \\   \pm \sqrt{3}/2, 
& -1/2\end{smallmatrix}\right)$ 
for  situations with the first and second wires detached, points $A_{1}$ and $A_{2}$.  Two chiral fixed points $\chi_{\pm}$, present in the broken time-reversal case, are 
$M_{1} (\alpha=\pm2\pi/3) =   \left( \begin{smallmatrix}
 -1/2  , &  \pm \sqrt{3}/2 \\   \mp \sqrt{3}/2, & -1/2\end{smallmatrix}\right)$; we have zero detached wires in this case.  
 
However, formally there is seventh fixed point, first discussed in \cite{Nayak1999} and denoted as $ D_{P}$ in 
\cite{Agarwal2010,Bellazzini2009,Das2008,Oshikawa2006}. It is given by 
$M_{1} (\alpha=\pi) =   \left( \begin{smallmatrix} -1  , &  0 \\   0, & -1 \end{smallmatrix}\right)$. 
From our Eq.\ (\ref{defGprime}) we have $G_{aa} =1$, $G_{bb}=4/3>1$, i.e.\ the 
value of one-terminal conductance $G_{bb}$ at the $D_{P}$ point is larger than the maximum value
of unity for the non-interacting situation. 
This enhanced value of conductance was interpreted as the signature of Andreev reflection and multiparticle 
scattering at the Y-junction. \cite{Nayak1999, Oshikawa2006}
We observe that for $D_{P}$ point one has $Tr(\bar M) =-1$, and it contradicts the charge conservation law both in 
quantum mechanical and classical formulation, at least for free fermions. Indeed, we should have $Tr(\bar M) \ge 0$  for any unitary $S$-matrix,
and for any doubly stochastic matrix, representing Kirchhoff's rule.     

One might argue, that our analysis, performed for free fermions, is inapplicable to the case of strongly interacting regime, where the importance of $D_{P}$ point was emphasized. It was predicted in \cite{Nayak1999,Oshikawa2006} that the fixed point $D_{P}$ becomes \emph{stable} at strong attraction between fermions, $g>3$, whereas it is unstable at weaker interaction, $g<3$ (cf.\ also \cite{Bellazzini2009}).  The existence of $D_{P}$ point, however, was not ruled out for any values of interaction and, particularly, in the free fermion limit, $g\to 1$. The inspection of formulas (10.29)--(10.31) in  \cite{Oshikawa2006} and Eq.\ (62) in \cite{Bellazzini2009} for the scaling dimension of leading perturbations confirms it. Therefore we have to conclude, that the discussed boundary for physical conductances was not so far detected by means of conventional bosonization. One possibility here, potentially favorable for the existence of $D_{P}$ point, is that this boundary changes with the Luttinger parameter $g$ and thus should be separately discussed for each value of $g$. Another possibility, supported by our analysis in \cite{Aristov2010,unpub},  is that the boundary determined for free fermions remains the same for any interaction strength. Particularly, any physical RG fixed point belongs to this boundary, while the above Fig.\ \ref{fig:deltoid}b  and the Fig.\ 5 in \cite{Aristov2010} depict the same parabola.  From this latter viewpoint, the discussed $D_{P}$ point lies outside the physical domain and is not connected with the inside of the body in Fig.\  \ref{fig:tetrahedron} by any RG flow. As was already mentioned in Sec.~\ref{sec:stability}, we assume the infinite non-interacting leads beyond the finite region of interaction, which setup provides the proper definition of $S$-matrix and vanishing vertex corrections to conductance.

Based on the above observations I make a following stronger conjecture. 
The two classes (\ref{M12}) do not represent 
any physical realization of $Y$-junction, except for six points listed above and corresponding to permutation matrices
in Eq. (\ref{permutmat}). These six points are the extreme points of the four-dimensional body, whose three-dimensional 
projections are shown in Fig.\   \ref{fig:tetrahedron}. All other physical realizations of the $Y$-junction lead to values 
$(x,y,z,\eta)$ lying strictly inside the four-dimensional unit sphere, Eqs.\ (\ref{restr-xyz}), (\ref{restrict}), whose surface 
is needed as a starting point in bosonization.  Thus the Abelian bosonization study of the $Y$-junction 
can be performed at six RG fixed points only,  at the vertices of the body in Fig.\   \ref{fig:tetrahedron}. 
If the RG fixed point occurs at the boundary of this body instead, then it is unsuitable for conventional 
bosonization analysis. It should be stressed again, that this conjecture is based on the present free-fermion analysis and on the results of perturbative two-loop RG treatment valid for relatively weak interaction. \cite{Aristov2010}  The case of strong interaction is to be discussed elsewhere. \cite{unpub}

\subsection{Larger-$N$ junctions}

The observations reported in this paper for the Y-junction, can be generalized for a larger number of semi-wires, $N>3$, attached to one spot. The $S$-matrix in this case will be given by a finite rotation matrix in $SU(N)$ group, and the squares of its entries form the unistochastic matrix $\bar M$.  The set of all such $\bar M$ lies within a set of bistochastic matrices, whose extreme points are $N!$ permutation matrices, according to the Birkhoff-von Neumann theorem. These permutation matrices   
describe the physical situations when some of the wires are detached from the central spot, whereas the other semi-wires form either ideal wires (in case of pairwise partial permutations) or chiral groups, involving three or larger number of semi-wires. For example, for $N=4$ (see also \cite{Bengtsson2005, Zyczkowski2003}) we obtain $24$  matrices, which correspond to 4 detached wires (one point in phase space), 2 detached wires and one full wire (six points), one detached wire and a chiral group of three semi-wires (eight points),  two full wires (three points),  and a chiral group of four semi-wires (3!=6 points). These points lie on an appropriate unit hyper-sphere and thus are suitable as fixed points in Abelian bosonization.  We note also that the particular case of fully symmetric $S$-matrix and fully symmetric interacting wires can be solved within TBA approach \cite{Egger2003}  for any $N$, because the high symmetry effectively reduces this case to $N=2$. The latter approach allows to consider the RG fixed point, corresponding to the perfectly transmitting junction ($z=-1/3$ in our above case $N=3$).

\begin{acknowledgments}
I am indebted to P. W\"olfle for numerous fruitful discussions. I am grateful to O. Yevtushenko, 
 V. Yu. Kachorovskii,  A. P. Dmitriev,   I. V. Gornyi,  D. G. Polyakov  for various
useful discussions. This work was supported by German-Israeli Foundation, 
DFG Center for Functional Nanostructures, Dynasty foundation, RFBR grants 09-02-00229, 11-02-00486. 
\end{acknowledgments}
 

\end{document}